\documentclass[manuscript]{emulateapj}

\usepackage{amsmath}
\usepackage{natbib}
\usepackage{graphicx}
\submitted{Accepted to The Astrophysical Journal}

\newcommand{\SO}{A0535+262}
\include{preamble}

\begin{document}

%% /*******************************************************************
%% ** The Header                                                     **
%% *******************************************************************/

\title{Search for Redshifted 2.2 MeV Neutron Capture Line from A0535+262 In Outburst}

\author{\c{S}irin \c{C}al\i\c{s}kan\altaffilmark{1},
        Emrah Kalemci\altaffilmark{1},
	Matthew G. Baring \altaffilmark{2},
        Steven E. Boggs\altaffilmark{3},
	Peter Kretschmar\altaffilmark{4}
}

\altaffiltext{1}{Sabanc\i\ University, Orhanl\i - Tuzla, \.Istanbul, 34956, TURKEY}

\altaffiltext{2}{Department of Physics and Astronomy, Rice University, 6100 Main, Houston, TX, 77005, USA}

\altaffiltext{3}{Space Sciences Laboratory, University of California, Berkeley, CA, 94720-7450, USA}

\altaffiltext{4}{ESA, European Space Astronomy Centre (ESAC), P.O. Box 78, 28691, Villanueva de la Ca\~{n}ada, Madrid, Spain.}

%% /*******************************************************************
%% ** The Abstract                                                   **
%% *******************************************************************/

\begin{abstract}

The Be/X-ray binary system \SO\ underwent a giant outburst in May-June 2005, 
followed by a dimmer outburst in August-September 2005. This increased 
intensity provided an opportunity to search for redshifted neutron-capture 
lines from the surface of the neutron star. If discovered, such lines would 
constrain the neutron star equation of state, providing the motivation of this 
search. The spectrometer (SPI) on board the \emph{INTEGRAL} satellite observed the 
dimmer outburst and provided the data for this research. We have not detected 
a line with enough significance, with the width-dependent upper limits on the 
broadened and redshifted neutron capture line in the range of 
(2 -- 11) $\times$ 10$^{-4}$ photons cm$^{-2}$ s$^{-1}$. To our knowledge, 
these are the strongest upper limits on the redshifted 2.2 MeV emission from 
an accreting neutron star. Our analysis of the transparency of the neutron star surface 
for 2.2 MeV photons shows that photons have a small but finite chance 
of leaving the atmosphere unscattered, which diminishes the possibility of
detection.

\end{abstract}

\keywords{nuclear reaction, nucleosynthesis, abundances -- stars: neutron -- X-rays: binaries 
-- stars: individual (A0535+262)}

%% /*******************************************************************
%% ** Introduction                                                   **
%% *******************************************************************/

\section{Introduction}\label{sec:intro}

Matter that is being accreted onto a neutron star has such large energies that 
it can undergo nuclear reactions by colliding with particles in the star's 
atmosphere. Nuclei that are heavier than Hydrogen will therefore break apart 
and generate free neutrons. The nuclear spallation process of $^{4}$He, 
discussed in detail in \citet{Bildsten93}(hereafter, \emph{BSW}), 
begins with the $^{4}$He atom 
entering the star's atmosphere with a very high kinetic energy. The atom 
undergoes many Coulomb collisions with atmospheric electrons and slows down.
The atom may hit an atmospheric proton and splits into a neutron and $^{3}$He. 
After this initial spallation, the $^{3}$He travels deeper into the atmosphere
 and thermalizes at a high column density. The thermal $^{3}$He can 
do one of three things: a) absorb a neutron (the most possible outcome),
 b) collide with a fast proton and be destroyed, or c) participate in a fusion 
reaction. The neutron that was liberated during the initial spallation 
undergoes elastic scatterings with atmospheric protons, thermalizes and drifts 
down due to gravity. Since most of the $^{3}$He atoms will capture the free 
neutrons, the rate of production of 2.2 MeV photons will depend on the amount 
of excess neutrons. Another restraining factor is the depth to which neutrons 
travel before recombining with protons. If recombination (and therefore 2.2 
MeV photon production) occurs deep in the atmosphere, the photon will have to 
undergo many Compton scatters before it can leave the atmosphere.
\citet{Bildsten93} 
have shown that for a neutron star with moderate magnetic field strength 
($\sim10^{9}$ G), a thermal neutron drifts approximately 1 Compton length 
before recombining and have calculated that only ~10\% of the 2.2 MeV photons 
will escape unscattered. 

These photons, emitted with an intrinsic energy \emph{E$_{0}$} from the atmosphere 
of a neutron star of radius \emph{R} and mass \emph{M}, experience a 
gravitational redshift for an observer at infinity to an energy \emph{E} 
given by

\[
\frac {R}{M} = \frac {2G}{c^{2}(1-\frac{E}{E_{0}})}
\]

\citep{Ozel03}. Measuring \emph{E/E$_{0}$} would allow a direct measurement 
of the gravitational redshift and place significant constraints on the 
underlying equation of state. Using the \emph{P}-$\rho$ relation for a given 
model, the equation of hydrostatic equilibrium can be numerically integrated 
to obtain the global mass-radius (\emph{M}-\emph{R}) relation for that equation
 of state, defining a single curve in the neutron star \emph{M}-\emph{R} 
diagram \citep{Lattimer04}. While the ratio \emph{M}/\emph{R} cannot identify 
the nuclear equation of state uniquely, it can rule out several models and 
place constraints on the remaining models. If the mass can be measured 
independently, using other means, then the nuclear equation of state can be
 uniquely identified.

When calculating the 2.2 MeV photon production mechanism, BSW have assumed
a moderate magnetic field  and that matter 
arrives radially onto the neutron star. The magnetic field of \SO\ has recently
been measured as $\sim 4 \times 10^{12}$G \citep{Caballero07}.
In such high magnetic fields, the ability of such line photons to escape
the neutron star is contingent upon the optical depths for Compton
scattering and also for magnetic pair creation being less than unity.
The Compton scattering constraint amounts to one on proximity of the
emission point to the upper layers of the stellar atmosphere; as
mentioned above, it was addressed in BSW, using non-magnetic Compton
scattering cross sections.  The pair production constraint is more
involved, and will be addressed at some length in the Discussion below.
The bottom line is that pair creation transparency of the magnetosphere
to 2.2 MeV photons will be achieved if their origin is at polar magnetic
colatitudes less than around 20 degrees {\it and} they are beamed within
around 15--20 degrees relative to their local surface magnetic field
vector.  These two requirements limit the expectation for visibility of
a 2.2 MeV emission line to small, but non-negligible probabilities.

The 2.2 MeV line is expected to be both redshifted and broadened, due to 
gravitational and relativistic effects \citep{Ozel03}.
Since we don't know a priori how much the line will be redshifted, we 
scanned the entire 1 MeV - 2.2 MeV region - 
wide enough to contain the redshifted line according to several different
equations of state \citep{Shapiro83} given in Table 1. The width of the 
redshifted 2.2 MeV line is also unknown, which led us to scan the 1 - 2 MeV 
region using different energy bands: 5keV, 11keV(${\Delta}$E/E$_{0}$ = 0.5\%),
 22 keV (${\Delta}$E/E$_{0}$ = 1\%), 44 keV 
(${\Delta}$E/E$_{0}$ = 2\%), 100 keV, 150 keV and 170 keV. We expect the
 line broadening to be minimal, since it is not a rapidly rotating neutron star 
with a pulsation period of 103 seconds \citep{Clark98}. 

\begin{table}[t]
\caption{\label{table:EOSs} Mass and Radius Values for Common Equations of 
State for Neutron Stars$^{a}$
 and the Expected Energies of the 2.2 MeV Line}
\begin{minipage}{\linewidth}
\begin{center}
\begin{tabular}{c|c|c|c} \hline \hline
Equation of State$^{b}$ & $M_{max}$/$M_{\odot}$ & R (km) & E (MeV) \\ \hline
$\pi$ & 1.5 & 7.5 & 1.41\\
R & 1.6 & 8.5 & 1.47 \\
BJ & 1.9 & 10.5 & 1.50 \\
TNI & 2.0 & 9 & 1.30 \\
TI & 2.0 & 12-15 & 1.56-1.72 \\
MF & 2.7 & 13.5 & 1.41 \\
\hline \hline
\end{tabular}\\
{$^{a}$ \cite{Shapiro83}

$^{b}$ $\pi$: Pion Condensate, R:Reid's, BJ: Bethe-Johnson, TNI: 3-Nucleon Interaction Approximation,
TI: Tensor Interaction, MF: Mean Field Approximation
}
\end{center}
\end{minipage}
\end{table}

The redshifted 2.2 MeV line from \SO was previously studied by 
\citet{Boggs06b}, using \emph{RHESSI} data. The \emph{RHESSI} data was gathered 
during \SO's giant outburst (May 18 - June 25), whereas we used data from a 
smaller outburst that took place approximately 3 months later \citep[see][for
a detailed discussion of \emph{INTEGRAL} and \emph{RXTE} analysis of the source]
{Caballero07}. The \emph{RHESSI} instrument uses Germanium crystals as 
detectors like SPI, but its area is smaller, and there is no shielding. It is 
specifically designed to study solar flares, and due to the close proximity of 
the Sun to this source during the giant outburst, while \emph{INTEGRAL} wasn't able 
to study this event, \emph{RHESSI} had an unique opportunity to observe the entire 
outburst. \citet{Boggs06b} subsequently calculated flux upper limits for \SO\, 
which we used for comparison.

%% /*******************************************************************
%% ** SPI Instrument and Analysis                                      **
%% *******************************************************************/

\section{SPI Instrument and Analysis}\label{sec:obs}

\emph{SPI} is the spectrometer on board the \emph{INTEGRAL} satellite that was launched in
October 2002. \emph{SPI} consists of 19 high purity Ge detectors, although only 17 are
currently operational since the loss of detectors 2 and 17 on December 6, 2003 
and July 18, 2004 respectively. \emph{SPI} has an energy range of 20 keV - 8 MeV with 
an energy resolution of 2.3 keV at 1.3 MeV \citep{Vedrenne03}. 
This high precision makes \emph{SPI} a good candidate for detecting narrow lines. 
The spectrometer uses coded aperture for imaging and is shielded by an 
anti-coincidence system that issues a veto flag for every photon that hits the 
detectors without passing through the mask.

The \emph{INTEGRAL} data we used was gathered between August 31 and September 2 2005 
(revolution 352). The good observing time was about 209 ksec, with a total of 
56 pointings. We tried four different background subtraction methods: two 
different empty fields, GEDSAT and GEDRATE. The first empty field we used was 
from revolution 220 (standard empty field in OSA, observations of Algol) and 
consisted of 103 pointings and 186 ksec. The second empty field we used was 
from revolutions 348 and 349 (observations of Cen-A, corresponding to August 
2005) closer by date to our observation, and consisted of 41 pointings and 
123 ksec. Although the latter one wasn't really ``an empty field'' 
in hard X-rays, no source was detected above 1 MeV. The GEDSAT and GEDRATE background 
models are integrated in the \emph{SPI} analysis software. GEDSAT assumes that the 
background level is proportional to the product of saturated detector trigger 
rate and live-time, whereas the GEDRATE background model assumes that the 
background level is proportional to the product of non-saturated detector 
trigger rate and live-time.

%Figure 1 used to be here.
\begin{figure}[t]
\centerline{\includegraphics[width=0.5\textwidth]{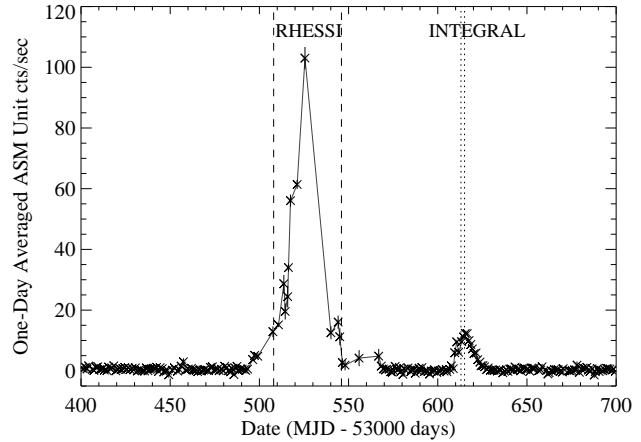}}
\caption{\label{fig:psds}
The light curve of \SO\ for the interval 30.01.2005 - 26.11.2005.
The spike around MJD 53530 represents the giant outburst and the smaller
 feature around
MJD 53620 represents the normal outburst in August - September 2005. 
}
\end{figure}

We used \emph{spiros} version 5.1 for data analysis, provided by ISDC. 
For each of the four background subtraction methods we accumulated data
for detector subsets: single events (SE), PSD events (PE), 
and double events (ME2). The single events represent the cases where a photon 
is detected by any one of the 17 are currently operational, but not tagged by 
the PSD. The PSD events represent the events which are analyzed by the Pulse 
Shape Discriminator. The double events (or multiple events with two detectors,
ME2) are defined as cases where groups of two detectors detect an event 
simultaneously, in which case the energy of the photon is given by the sum of 
the two detectors. The reason we have different detector subsets is as follows.
We do not want to use only SE+PE events (which is usually done for analysis of
low energy data) because there is significant area at ME2 above 1 MeV. 
Moreover, for SE there is a significant background noise in the 1.4 - 1.7 MeV 
region \citep{Weidenspointner03} which is artificial. These background
features are diminished when double events and/or PSD events are used. 
Therefore except in this region we merged PE, SE, and ME2, while within the region
of the artificial 
background component we used PE+ME2. This was done by using a modified 
\emph{spi\_obs\_hist}, and renormalizing the effective area after the fluxes 
are obtained. The rest of the analysis has been done using standard OSA 
software.

In order to create the spectrum of \SO\, we did imaging in 5keV steps and 
plotted the flux (photons cm$^{-2}$ sec$^{-1}$ keV$^{-1}$)
against energy. We used IDL to fit Gaussian curves of given energy and FWHM 
and extracted the errors of the flux, given by the area under the Gaussian, to
 calculate the 3 $\sigma$ flux upper limits. 

%% /*******************************************************************
%% ** Results                                                        **
%% *******************************************************************/

\section{Results}\label{sec:results}

Two spectra of \SO\ are given in Figures 2 and 3, for different background 
subtraction methods and different detector subsets.
The significance of \SO\ did not exceed a few $\sigma$ in any of the imaging 
analyses, therefore we couldn't detect a line. However, 
we calculated upper limits for the flux for different background 
subtraction methods and line widths. In order to do this, we fit the residual 
spectrum to Gaussian lines with a constant background in 5 keV steps.
The errors of these fits provided us with the flux upper limits. When fitting 
Gaussians with large FWHM values (such as 44 keV or 110 keV), the low and high 
energy limits of the spectrum gave unusually high error values due to the lack 
of data below 1 MeV and above 2.2 MeV; these were neglected.
We chose to use the same FWHM values as \citet{Boggs06b} in order to make a 
meaningful comparison of the \emph{RHESSI} and \emph{SPI} instruments. The 
line widths range from 11 keV FWHM (${\Delta}$E/E$_{0}$ = 0.5\%) to 110 keV 
FWHM (${\Delta}$E/E$_{0}$ = 5.0\%) since we cannot calculate the exact line 
broadening, which depends on the geometry of the mass accretion and emission.
Three energy dependent flux upper limit plots are shown in Figures 4, 5 and 6 
and the upper limits for all background subtraction methods and line widths 
are given in Table 2 along with the \emph{RHESSI} upper limits for comparison
 \citep{Boggs06b}.

%Figures 2 - 6 used to be here.
\begin{figure}[t]
\centerline{\includegraphics[width=0.5\textwidth]{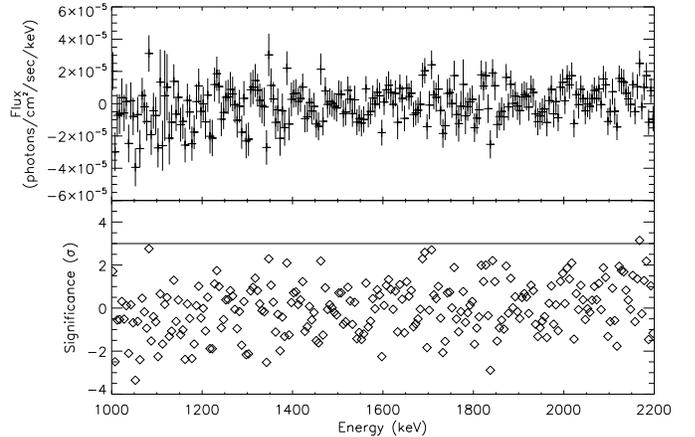}}
\caption{\label{fig:psds}
The upper panel shows the background subtracted spectrum of \SO\, using 
revolution 220 as flat field. Only single events and PSD
events were used in this analysis. The lower panel shows the significance of 
\SO\ in the 1-2.2 MeV region. The 3$\sigma$ threshold is rarely crossed (not
verified by other background methods and energy bands), which prompted us to 
calculate upper limits for the flux.
}
\end{figure}

\begin{figure}[t]
\centerline{\includegraphics[width=0.5\textwidth]{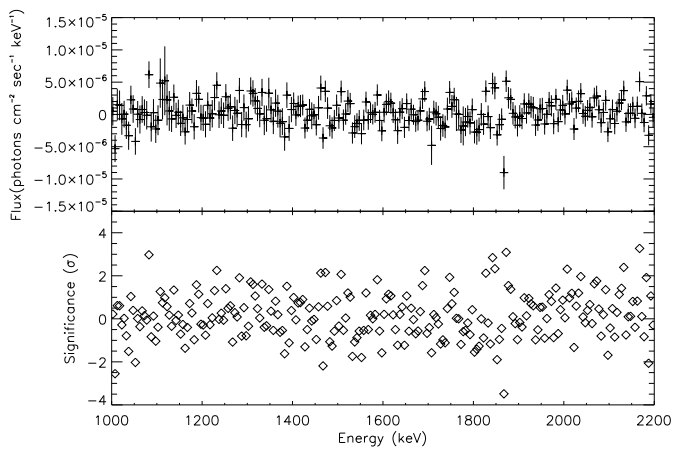}}
\caption{\label{fig:psds}
The upper panel shows the background subtracted spectrum of \SO\, corrected for
 the \emph{SPI} effective area and using GEDSAT for background subtraction. 
Single events, double events and PSD
events were used in this analysis. The lower panel shows the significance of 
\SO\ in the 1-2.2 MeV region.
}
\end{figure}

\begin{figure}[t]
\centerline{\includegraphics[width=0.5\textwidth]{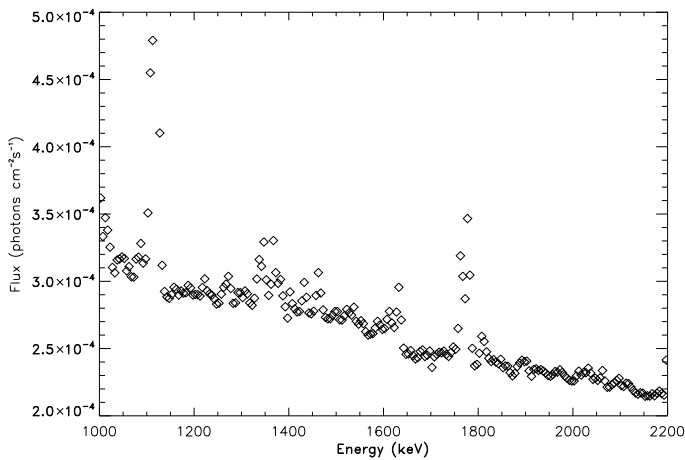}}
\caption{\label{fig:psds}
The 3 $\sigma$ upper limits of \SO\ using single and PSD events, Revolution 
220 as flat field and fitting a Gaussian with 11keV FWHM (${\Delta}$E/E$_{0}$ 
= 0.5\%). The peak around 1100 keV is due to electron pair production line. 
The 1400-1700 keV region is dominated by instrumental background lines, 
therefore only PSD events were taken into account in that region. The peak 
around 1800 keV is another instrumental background line 
\citep{Weidenspointner03}.
}
\end{figure}

\begin{figure}[t]
\centerline{\includegraphics[width=0.5\textwidth]{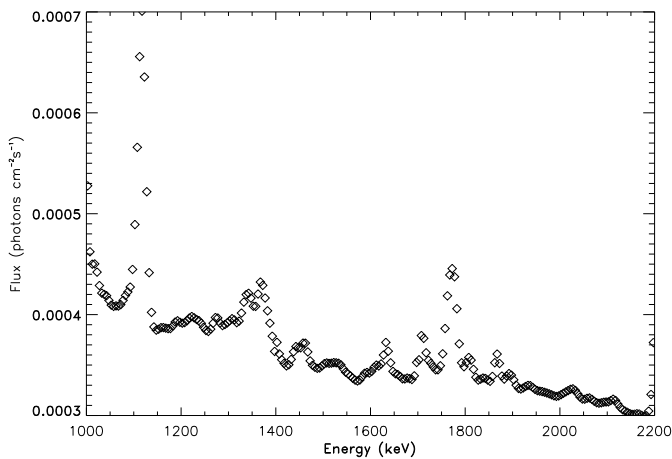}}
\caption{\label{fig:psds}
The 3 $\sigma$ upper limits of \SO\ using single, multiple and PSD events, 
GEDSAT for background 
subtraction and fitting a Gaussian with 22keV FWHM.
}
\end{figure}

\begin{figure}[t]
\centerline{\includegraphics[width=0.5\textwidth]{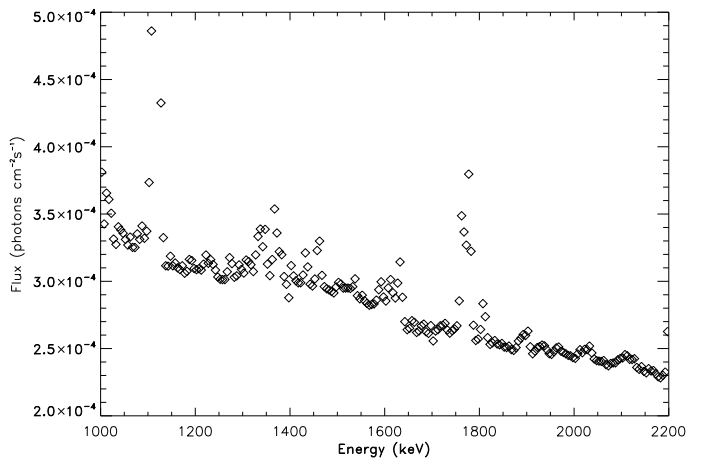}}
\caption{\label{fig:psds}
The 3 $\sigma$ upper limits of \SO\ using single and PSD events, GEDSAT as
 flat field and fitting a Gaussian with 11keV FWHM.
}
\end{figure}
%Table 2 used to be here.

\begin{table*}[t]
\caption{\label{table:upperlimit} Redshifted 2.2 MeV Line, 3 $\sigma$ Upper 
Limits, 1-2.2 MeV}
\begin{minipage}{\linewidth}
%\renewcommand{\thefootnote}{\thempfootnote}
%\scriptsize
\begin{center}
\begin{tabular}{c|c|c|c|c|c|c} \hline \hline
${\Delta}$E/E$_{0}$ (FWHM) & RHESSI Flux$^{a}$ & Rev. 220 Flux$^{b}$ & Rev 348/349$^{b}$ Flux & GEDSAT 
Flux$^{b}$ & GEDRATE Flux$^{b}$ & GEDSAT Flux$^{c}$\\ 
(\%) & (photons cm$^{-2}$ s$^{-1}$) & (photons cm$^{-2}$ s$^{-1}$)& (photons cm$^{-2}$ s$^{-1}$) &
 (photons cm$^{-2}$ s$^{-1}$) & (photons cm$^{-2}$ s$^{-1}$) & (photons cm$^{-2}$ s$^{-1}$)\\ \hline
0.5 & 4.0 $\times$ 10$^{-4}$ & 2-3 $\times$ 10$^{-4}$ & 2-3 $\times$ 10$^{-4}$ & 2-3 $\times$ 10$^{-4}$  &
 2-3 $\times$ 10$^{-4}$ & 2-3 $\times$ 10$^{-4}$\\
1.0 & 5.0 $\times$ 10$^{-4}$ & 3-4 $\times$ 10$^{-4}$ & 3-4 $\times$ 10$^{-4}$ & 3-5 $\times$ 10$^{-4}$  &
 3-5 $\times$ 10$^{-4}$ & 3-4 $\times$ 10$^{-4}$\\
2.0 & 6.4 $\times$ 10$^{-4}$ & 4-6 $\times$ 10$^{-4}$ & 4-6 $\times$ 10$^{-4}$ & 5-7 $\times$ 10$^{-4}$  &
 5-7 $\times$ 10$^{-4}$ & 4-6 $\times$ 10$^{-4}$\\
5.0 & 10.5 $\times$ 10$^{-4}$ & 7-10 $\times$ 10$^{-4}$ & 7-10 $\times$ 10$^{-4}$ & 8-11 $\times$ 10$^{-4}$
  & 8-11 $\times$ 10$^{-4}$ & 7-10 $\times$ 10$^{-4}$\\
\hline \hline
\end{tabular}\\
{$^{a}$ \cite{Boggs06b}

$^{b}$Using single and PSD events only.

$^{c}$Using single, double and PSD events.
}
\end{center}
\end{minipage}
\end{table*}

%% /*******************************************************************
%% ** Discussion                                                        **
%% *******************************************************************/

\section{Discussion}\label{sec:discussion}

Despite the excellent energy resolution and relatively high area of \emph{SPI} for 
high energy photons, the redshifted 2.2 MeV line from \SO\ was not detected
significantly. However, we were able to obtain stronger
constraints in the flux than that of \emph{RHESSI} instrument. 
The non detection is not surprising, as the transparency of the neutron star, 
detailed below,
is greatly diminished for high magnetic fields, reducing the yield of 2.2 MeV photons.

The visibility of any atmospheric 2.2 MeV neutron capture line is
subject to potential attenuation in the neutron star magnetosphere.  The
leading process for such absorption is magnetic pair production,
$\gamma\to e^+e^-$, the process that underpins standard radio and polar
cap gamma-ray pulsar models (e.g. \citet{Sturrock71}, \citet{Daugherty96}).  
This mechanism is the highest order photon absorption
interaction in strong-field QED, and is prolific at the field strengths
inferred for A0535+262.  It has a threshold (for the $\parallel$
polarization state) of $2 m_ec^2/\sin\theta_{kB}$, where $\theta_{kB}$
is the angle the photon momentum vector makes to the local field line,
above which the interaction rate precipitously rises (e.g. \citet{Daugherty83}, 
\citet{Baring08}. 
At energies only modestly above threshold,
the attenuation lengths in local fields greater than around $10^{12}$
Gauss are considerably less than 1cm. Accordingly, once the threshold is
crossed, 2.2 MeV photons can be expected to be destroyed in A0535+262,
provided this occurs at altitudes less than around one stellar radius
above the surface.  This accessibility of the threshold defines the
broad criterion for transparency versus opacity of the magnetosphere for
the line photons, the outcome of which is critically dependent on
geometry.  Attenuation phase space can readily be assessed as functions
of the surface polar field, the surface emission locale, and the
emission direction of the photons.  Such a detailed analysis was offered
by \citet{Baring01} at much higher energies that are germane to
pulsar and magnetar continuum emission.  It should be remarked that
familiar two photon pair creation is negligible in comparison.

The transparency of the magnetosphere for 2.2 MeV line photons was investigated 
here principally via the salient threshold criterion, to quickly ascertain the 
pertinent approximate phase space.  Our analysis was performed in flat spacetime, 
sufficient for a general indication of transparency, for which rather simple 
geometrical relations for photon transport through the magnetosphere can be 
prescribed, along the lines of the approach that \citet{Baring07} used 
for resonant Compton upscattering in magnetars.  The involved mathematical details 
of this analysis will be omitted here, for the sake of brevity, and only the 
principal conclusions outlined.  Photons at 2.2 MeV that are emitted at the 
atmospheric surface are below magnetic pair creation threshold when they initially 
have $\sin\theta_{kB} 2 m_ec^2/(2.2 MeV) \Rightarrow \theta_{kB}\lesssim 27^\circ$.  
Note that at the field strengths of $\lesssim 4\times 10^{12}$ Gauss appropriate 
for A0535+262, the pair threshold for the $\perp$ photon polarization state is 
extremely close in energy to that for the $\parallel$ state.  For equatorial 
emission locales, where the radius of field curvature is comparatively small, 
even photons that are initially beamed close to field lines rapidly acquire 
sufficient angles with respect to {\bf B} to precipitate pair attenuation. 
Hence the magnetosphere is necessarily opaque to $\gamma\to e^+e^-$ for emission 
colatitudes $\theta_e$ generally greater than around $30^\circ$. Near the pole, 
if $\theta_{kB}\lesssim 27^\circ$ initially, transport of photons to significant 
altitudes is required to access pair threshold. There, the field line curvature 
radius is $\rho_c\approx 4R_{ns}/(3\sin\theta_e)$ for a neutron star radius of 
$R_{ns}$. Accordingly, photons that initially are beamed close to the field, 
acquire an angle of $\theta_{kB}\gtrsim 27^\circ$ when reaching an altitude 
$h\sim 4R_{ns}\sin 27^\circ/(3\sin\theta_e)$.  This altitude must generally be 
more than around a stellar radius for the field decline to render the pair creation 
rate insufficient to sustain photon attenuation, as opposed to free escape (see, 
e.g. \citet{Baring01} for a detailed examination). To summarize our analysis, 
here we find that pair creation transparency of the magnetosphere to 2.2 MeV photons 
will occur if their origin is at polar magnetic colatitudes $\theta_e \lesssim 20^\circ$ 
{\it and} they are initially beamed within around 15--20 degrees relative to their local 
surface magnetic field vector. These two requirements limit the expectation for 
visibility of any 2.2 MeV emission line to modest or small, but non-negligible, 
surface areas and solid angles.

With a more powerful tool for background
rejection and a dramatical improvement in sensitivity (broad line sensitivity of
1.2 $\times$ 10$^{-6}$ photons cm$^{-2}$ s$^{-1}$ and a spectral resolution of 0.2 -- 1 $\%$)
\citep{Boggs06_act}, the Advanced Compton Telescope could be a good opportunity 
to follow up on the search for the broadened and redshifted 2.2 MeV line. 

%% /*******************************************************************
%% ** Acknowledgments                                               **
%% *******************************************************************/

\acknowledgments

This project was supported by the European Commission through a FP6 
Marie-Curie International Reintegration Grant (INDAM, MIRG-CT-2005-017203). 
E.K. acknowledges support of T\"UB\.ITAK Career Development Award, and also
Turkish Academy of Sciences Young and Successful Scientist Award (T\"UBA, GEB\.IP). 
This project 
is supported by NASA Grant NNX08AC94G. E.K. and \c{S}.\c{C}. also acknowledge 
support from FP6 Marie Curie Actions Transfer of Knowledge (ASTRONS, 
MTKD-CT-2006-042703). \c{S}.\c{C}. thanks Feryal \"Ozel and 
Dimitrios Psaltis for useful discussions. R.E. Rothschild provided the Cen A 
observations.

\end{document}